\documentclass[journal]{IEEEtran}
\usepackage{amsmath,amsfonts}
\usepackage{algorithmic}
\usepackage{algorithm}
\usepackage{array}
\usepackage[caption=false,font=normalsize,labelfont=scriptsize,textfont=scriptsize]{subfig}
\usepackage{textcomp}
\usepackage{stfloats}
\usepackage{url}
\usepackage{verbatim}
\usepackage{graphicx}
\usepackage{cite}
\usepackage{amssymb}
\usepackage{float}
\usepackage{color}

\allowdisplaybreaks
\begin{document}
\title{\huge Enhanced Channel Estimation for Flexible Intelligent Metasurface-Aided Communication Systems}
\author{Jinyue Jiang, Jiancheng An,~\IEEEmembership{Senior Member,~IEEE,} Lu Gan,~\IEEEmembership{Member,~IEEE,} \\Naofal Al-Dhahir,~\IEEEmembership{Fellow,~IEEE,} Arumugam Nallanathan,~\IEEEmembership{Fellow,~IEEE,} and Zhu Han,~\IEEEmembership{Fellow,~IEEE}
\vspace{-1em}
\thanks{This work was partially supported by National Natural Science Foundation of China 62471096. The work of N. Al-Dhahir was supported by Erik Jonsson Distinguished Professorship. \textit{(Corresponding author: Jiancheng An.)}}
\thanks{J. Jiang and L. Gan are with the School of Information and Communication Engineering, University of Electronic Science and Technology of China (UESTC), Chengdu, Sichuan, 611731, China. L. Gan is also with the Yibin Institute of UESTC, Yibin 644000, China (e-mail: jinyue\_jiang@std.uestc.edu.cn; ganlu@uestc.edu.cn).}
\thanks{J. An is with the School of Electrical and Electronics Engineering, Nanyang Technological University, Singapore 639798 (e-mail: jiancheng\_an@163.com).}
\thanks{N. Al-Dhahir is with the Department of Electrical and Computer Engineering, The University of Texas at Dallas, Richardson, TX 75080 USA (e-mail: aldhahir@utdallas.edu).}
\thanks{A. Nallanathan is with the School of Electronic Engineering and Computer Science, Queen Mary University of London, London, U.K., and also with the Department of Electronic Engineering, Kyung Hee University, Yongin-si, Gyeonggi-do 17104, Korea (e-mail: a.nallanathan@qmul.ac.uk).}
\thanks{Z. Han is with the Department of Electrical and Computer Engineering at the University of Houston, Houston, TX 77004 USA, and also with the Department of Computer Science and Engineering, Kyung Hee University, Seoul, South Korea, 446-701 (e-mail: hanzhu22@gmail.com).}}
\markboth{DRAFT}{DRAFT}
\maketitle
\vspace{-5em}
\begin{abstract}
Flexible intelligent metasurface (FIM) has recently received considerable interest due to its advantage in realizing a better channel condition by dynamically morphing its surface shape. An FIM consists of multiple elements deposited on a flexible substrate. These elements can not only transmit signals, but also adapt their displacements in a direction perpendicular to the FIM surface via an attached controller. In this paper, we consider the channel estimation problem for the uplink of an FIM-enhanced communication system via customizing the orthogonal matching pursuit (OMP) method. Specifically, we formulate an optimization problem of minimizing the column coherence of the measurement matrix by optimizing the FIM's surface shape, subject to the morphing range constraint. Based on the estimated direction of arrival (DOA) and channel gain, we further investigate the signal-to-noise ratio (SNR) improvement in the FIM-enhanced downlink multiple-input single-output (MISO) system. Numerical results demonstrate that an FIM significantly outperforms a conventional rigid uniform planar array (UPA), thereby showing that FIM can substantially improve channel estimation accuracy and achieve SNR improvement, even when using estimated channel parameters.
\end{abstract}

\vspace{-1em}
\section{Introduction}
\IEEEPARstart{T}{he} sixth-generation (6G) wireless networks impose unprecedented requirements on data rate and reliability, which requires developing new communication technologies\cite{9040264, WC_2024_An_Codebook}. Recently, advanced metasurface antennas have been designed to employ feeds to inject surface-propagating electromagnetic waves and leverage the radiation properties of metallic patches for precise electromagnetic control\cite{zhang2024target}. In the past few years, metasurfaces have been extensively studied to demonstrate their effectiveness in enhancing beamforming and signal processing capabilities in communication and sensing systems\cite{yuan2025reconfigurable}.

Although multi-antenna arrays and metasurfaces provide significant performance gains \cite{10158690, TCOM_2024_Yu_Environment}, these technologies rely on a rigid array substrate, unable to take full advantage of dynamic wireless channels that change continuously in the spatial domain \cite{zheng2023flexible}. Specifically, in a multipath environment, signals from different paths are superimposed with varying amplitudes and phases, causing the received signal power to fluctuate across the receiving array. Flexible intelligent metasurfaces (FIMs), including foldable metasurfaces \cite{FIM_folding}, stretchable metasurfaces \cite{FIM_stretchable}, etc., can further enhance the spatial degrees of freedom of communication systems based on their deformation mechanisms. However, the deformation mechanisms of these metasurfaces currently lack programmably controllable structures.

Recently, the authors of \cite{ni2021dynamically} proposed a programmable three-dimensional flexible intelligent metasurface (FIM), which has the ability to morph its surface shape quickly and accurately by leveraging advanced EM actuation technology\footnote{Please refer to the video on https://www.eurekalert.org/multimedia/950133, which visually presents the real-time surface shape morphing and retention capabilities of FIM. The FIM in \cite{ni2021dynamically} can achieve a displacement resolution of up to 0.006 millimeters, which can be regarded as continuous morphing.}. This structure makes it possible to effectively implement MIMO with flexible antenna arrays \cite{an2024emerging}. Specifically, an FIM is composed of multiple elements connected to each other by flexible material (e.g., filamentary metals\cite{ni2021dynamically}) or flexible structure (e.g., hydrogels, nanocomposites\cite{niu2021reconfigurable}, cellulosic material\cite{zhang2023self}), allowing for more adaptable deployment on flexible objects, such as curtains. Each element can transmit signals, while adapting its displacement in a direction perpendicular to the surface with the aid of controllers. This characteristic renders FIM to be a promising candidate for transceiver arrays in communication systems. Specifically, by morphing the FIM surface shape such that multiple signal copies from different paths are coherently superimposed across the morphed surface, the channel gain can be significantly enhanced. 

In \cite{an2025flexibleMISO, an2025flexibleMIMO}, the authors investigated the performance gains of FIM in terms of energy efficiency and channel capacity. \cite{jiang2023bivariate} investigated the issues of channel estimation and data transmission in a reconfigurable intelligent surface (RIS)-aided multi-user multiple-input single-output orthogonal frequency division multiplexing (MU-MISO-OFDM) system, and minimizes the column coherence of the measurement matrix by optimizing pilot positions and pilot power. However, the effective channel estimation algorithm for FIM-aided systems as well as the surface shape morphing algorithm for enhancing channel estimation accuracy has not been examined yet in the open literature.

Motivated by this observation, in this paper, we utilize an FIM as a receive antenna array to enhance uplink channel estimation in a single-user scenario. Specifically, this paper employs the orthogonal matching pursuit (OMP) algorithm for wireless channel estimation. Unlike existing benchmark methods, this paper focuses on minimizing the column coherence of the angular-domain measurement matrix by morphing the surface shape of the FIM, thereby improving the FIM channel estimation accuracy of the OMP algorithm. Furthermore, we utilize the estimated direction of arrival (DOA) and channel gain to optimize FIM-enhanced downlink multiple-input single-output (MISO) communication and analyze the impact of channel parameters' errors on the signal-to-noise ratio (SNR) gain achieved by FIM. Numerical results demonstrate the effectiveness of FIM in improving the accuracy of estimating the DOA and path gain. Moreover, FIM can significantly improve downlink SNR by utilizing estimated channel parameters.

\textit{Notations:} 
$(\cdot)^T$, $(\cdot)^*$, and $(\cdot)^H$ denote the transpose, conjugate and conjugate transpose, respectively. $\boldsymbol{I}_N$ denotes the $N\times N$ identity matrix. The 2-norm of vector $\boldsymbol{a}$ is denoted by $\Arrowvert\boldsymbol{a}\Arrowvert_2$. The amplitude of complex number $a$ is denoted by $\arrowvert a \arrowvert$. $\odot$ represents the Hadamard product. $\boldsymbol{A} \succ 0$ refers to a positive definite matrix $\boldsymbol{A}$. $\Im$ denotes the imaginary part of a complex number. $\mathbb{E}[\cdot]$ denotes the statistical expectation operator. $\mathbb{R}$ and $\mathbb{C}$ denote the set of real and complex numbers, respectively.

\section{FIM-Aided Uplink SIMO System Model}
\begin{figure}[!t]
\centering
\includegraphics[width=\linewidth]{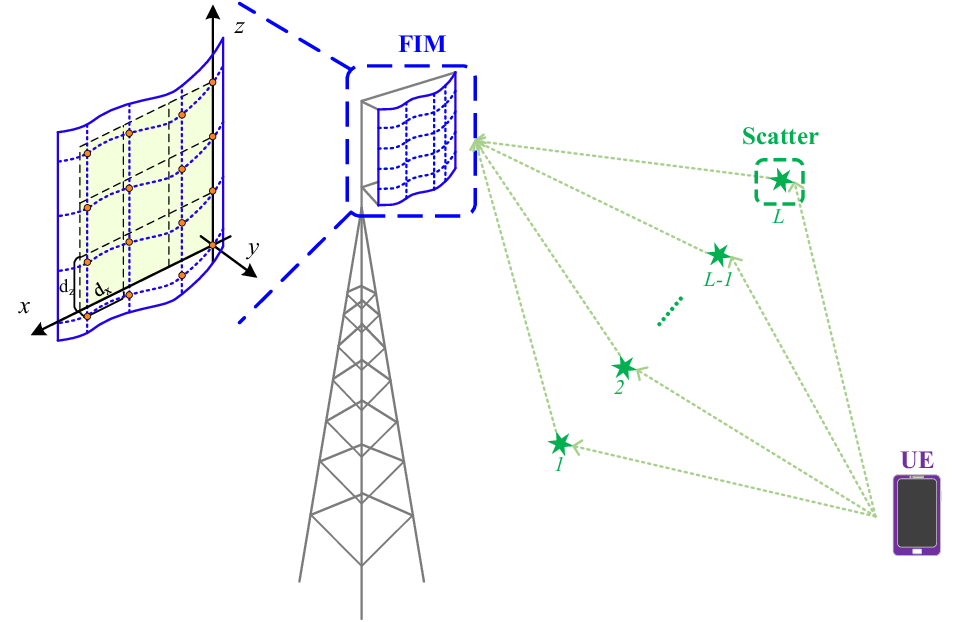}
\caption{An FIM-enhanced SIMO system, where the surface shape of the FIM can be morphed to enhance the channel estimation accuracy.}
\label{fig_1}
\end{figure}

As shown in Fig. \ref{fig_1}, we consider an FIM-enhanced uplink single-input multiple-output (SIMO) system with $N$ elements at the receiver to perform channel estimation in the cell. The position of the $n$-th element can be represented by its 3D Cartesian coordinates $\boldsymbol{t}_n=[x_n,y_n,z_n]^T \in \mathbb{R}^3$, $n=1,2,\ldots,N$. Without loss of generality, we assume that the projection of all FIM meta-atoms on the $x$-$z$ plane is modeled as a rigid uniform planar array (UPA). The element spacings in the $x$-axis and $z$-axis are denoted by $d_x$ and $d_z$, respectively. According to \cite{ni2021dynamically}, the projection of the element coordinates on the $x$-$z$ plane remains constant as the FIM morphs its surface shape. Moreover, we assume that the deformation of all elements along the $y$-axis is restricted to $\mathcal{B}=\{y_n|-b \leq y_n \leq b, n=1,2,\ldots,N\}$, where $b$ characterizes the maximum morphing range of the FIM. As a result, $\boldsymbol{y}=[y_1, y_2,\ldots, y_N]^T \in \mathbb{R}^{N}$ characterizes the FIM's surface shape. 

Furthermore, we consider quasi-static flat fading channels and use the multipath propagation model to characterize the channel spanning from the user equipment (UE) to the base station (BS). Let $L$ denote the number of propagation paths. Specifically, the wireless channel is modeled as
\begin{equation}
\label{deqn_ex1}
\boldsymbol{h}(\boldsymbol{y})=\sum_{l=1}^L\xi_l\boldsymbol{g}_l(\boldsymbol{y})\in \mathbb{C}^N,
\end{equation}
where $\xi_l$ denotes the complex gain of the $l$-th path from the UE to the FIM, following a circularly symmetric complex Gaussian (CSCG) distribution $\xi_l\sim\mathcal{CN}(0,\gamma^2/L)$, with $\gamma^2$ denoting the average channel gain. $\boldsymbol{g}_l(\boldsymbol{y})$ denotes the steering vector of the $l$-th propagation path\cite{heath2016overview}, which is given by

\begin{equation}
\boldsymbol{g}_l(\boldsymbol{y})=[e^{j\frac{2\pi}{\lambda}\rho_1^l},e^{j\frac{2\pi}{\lambda}\rho_2^l},\ldots,e^{j\frac{2\pi}{\lambda}\rho_N^{l}}]^T \in \mathbb{C}^{N},
\end{equation}
where $\rho_n^l=x_n\sin{\theta^l}\cos{\phi^l}+y_n\sin{\theta^l}\sin{\phi^l}+z_n\cos{\theta^l}$ is the propagation distance difference between the $n$-th meta-atom and the origin of the coordinate system projected on the direction of the $l$-th path, while $\theta^l\in[0,\pi)$ and $\phi^l\in[0,\pi)$ denote the elevation and azimuth angles of the $l$-th propagation path's DOA, respectively. $\lambda$ denotes the radio wavelength. Note that in contrast to conventional SIMO communications, the wireless channel $\boldsymbol{h}(\boldsymbol{y})$ is a function of the FIM surface shape $\boldsymbol{y}$.

Let $s_\text{up}\in \mathbb{C}$ represent the pilot signal sent by the UE. Therefore, the received signal $\boldsymbol{f}_{\rm up} \in \mathbb{C}^N$ is given by 

\begin{equation}
\label{equ_up}
\boldsymbol{f}_{\rm up}(\boldsymbol{y})=\boldsymbol{h}(\boldsymbol{y})s_{\rm up}+\boldsymbol{\eta}_{\rm up},
\end{equation}
where $\boldsymbol{\eta}_{\rm up}\sim\mathcal{CN}(\boldsymbol{0},\sigma^2_{\rm up}\boldsymbol{I})$ is the additive white Gaussian noise (AWGN) at the BS with $\sigma^2_{\rm up}$ denoting the average noise power at the FIM.

\section{Enhanced Channel Estimation via FIM Surface Shape Morphing} \label{ES-CSI}
\subsection{Compressed Sensing-based Channel Estimation Method}
In future 6G wireless networks, the introduction of high-frequency large-scale antenna arrays will typically result in channel sparsity in the angular domain. In this section, we investigate uplink channel estimation in FIM-aided communication systems by utilizing this sparsity property. In contrast to directly estimating $\boldsymbol{h}(\boldsymbol{y})$, we aim to estimate the channel parameters, including multipath complex gain $\{\xi^l\}$ and DOA parameters $\{ \theta^l, \phi^l \}$. To this end, (\ref{equ_up}) can be rewritten as:
\begin{equation} \label{sparse_model}
\boldsymbol{f}_{\rm up}(\boldsymbol{y})=\boldsymbol{G(\boldsymbol{y})}\boldsymbol{\xi}s_{\rm up}+\boldsymbol{\eta}_{\rm up}=\boldsymbol{\Psi}(\boldsymbol{y})\boldsymbol{\beta}s_{\rm up}+\boldsymbol{\eta}_{\rm up}, 
\end{equation}
where $\boldsymbol{\xi}=[\xi_1,\xi_2,\ldots,\xi_L]^T\in \mathbb{C}^{L}$, $\boldsymbol{G}(\boldsymbol{y})=[\boldsymbol{g}_1,\boldsymbol{g}_2,\ldots,\boldsymbol{g}_L]\in \mathbb{C}^{N\times L}$. Furthermore, $\boldsymbol{\beta} \in \mathbb{C}^{M}$ represents $M$-dimensional sparse vector with a sparsity level of $L$, with $\boldsymbol{\xi}$ being the non-zero gain associated with the support set and $M$ representing the dictionary size. The measurement matrix $\boldsymbol{\Psi}(\boldsymbol{y})$ is constructed as
\begin{equation}
 \boldsymbol{\Psi}(\boldsymbol{y})=[\boldsymbol{\psi}_1(\boldsymbol{y}),\boldsymbol{\psi}_2(\boldsymbol{y}),\ldots,\boldsymbol{\psi}_{M}(\boldsymbol{y})]\in \mathbb{C}^{N\times {M}},
\end{equation}
where $\boldsymbol{\psi}_m(\boldsymbol{y})$ is defined by \vspace{-0.5em}

\begin{equation}
 \boldsymbol{\psi}_m(\boldsymbol{y})=[e^{j\frac{2\pi}{\lambda}\hat{\rho}_1^m},e^{j\frac{2\pi}{\lambda}\hat{\rho}_2^m},\ldots,e^{j\frac{2\pi}{\lambda}\hat{\rho}_N^m}]^T \in \mathbb{C}^{N},
\end{equation}
with $\hat{\rho}_n^m=x_n\varphi_x^m+y_n\varphi_y^m+z_n\varphi_z^m$. The angle parameters $\varphi_x^m=\sin{\theta}^m\cos{\phi}^m$ and $\varphi_z^m=\cos{\theta}^m$ are obtained by uniformly quantizing the interval $[-1, 1]$ using $M_x$ and $M_z$ points, respectively, satisfying $M_xM_z = M$. More specifically, we have $\varphi_x^m=-1+(2i-1)/M_x, \varphi_z^m=-1+(2j-1)/M_z$, where $i$ and $j$ are given to satisfy $m=(i-1)M_z+j$, with $i=1,2,\dots,M_x$, $j=1,2,\dots,M_z$.

Accordingly, the corresponding electrical angle $\varphi_y^m=\sin{\theta}^m\sin{\phi}^m$ in the $y$-direction can be derived as $\varphi_y^m=\sqrt{1-(\varphi_x^m)^2-(\varphi_z^m)^2}$.

Furthermore, we consider collecting the pilot signal over multiple consecutive time slots to enhance the channel estimation accuracy. By collecting the total received signal in (\ref{sparse_model}) over $N_s$ time slots, we arrive at
\begin{equation}
\label{equ_up_all}
\tilde{\boldsymbol{f}}_{\rm up}=\tilde{\boldsymbol{\Psi}}\boldsymbol{\beta}s_{\rm up}+\tilde{\boldsymbol{\eta}}_{\rm up},
\end{equation}
where $\tilde{\boldsymbol{f}}_{\rm up}=[\boldsymbol{f}^T_{\rm up}(\boldsymbol{y}^1),\ldots,\boldsymbol{f}^T_{\rm up}(\boldsymbol{y}^{N_s})]^T\in \mathbb{C}^{N_sN}$, $\tilde{\boldsymbol{\Psi}}=[\boldsymbol{\Psi}^T(\boldsymbol{y}^1),\ldots,\boldsymbol{\Psi}^T(\boldsymbol{y}^{N_s})]^T \in \mathbb{C}^{N_sN \times M}$, $\tilde{\boldsymbol{\eta}}_{\rm up}=[(\boldsymbol{\eta}_{\rm up}^1)^T,\ldots,(\boldsymbol{\eta}_{\rm up}^{N_s})^T]^T\in \mathbb{C}^{N_sN}$. Note that (\ref{equ_up_all}) is a sparse recovery problem, which can be solved by using compressed sensing-based methods, e.g., the OMP algorithm \cite{8240645}. 

\subsection{FIM Surface-Shape Morphing for Optimizing $\tilde{\boldsymbol{\Psi}}$} \label{DFP_method}
In contrast to conventional rigid antenna arrays, the FIM surface shapes over $N_s$ training slots can be further morphed to minimize the column coherence of the measurement matrix $\tilde{\boldsymbol{\Psi}}$, thus implicitly reducing the restricted isometric constant and achieving better estimation results. Specifically, by constructing an objective function to minimize the sum of coherence coefficients between arbitrary pairs of columns in $\tilde{\boldsymbol{\Psi}}$, we introduce an objective function as \vspace{-0.5em}

\begin{equation}
\gamma(\tilde{\boldsymbol{y}})=\sum_{m=1}^M\sum_{v=m+1}^M\arrowvert 
\boldsymbol{\phi}^H_m \boldsymbol{\phi}_v \arrowvert,
\end{equation}
where $\boldsymbol{\phi}_m$ denotes the $m$-th column of the measurement matrix $\tilde{\boldsymbol{\Psi}}$, $\tilde{\boldsymbol{y}}=[(\boldsymbol{y}^1)^T, (\boldsymbol{y}^2)^T, \dots, (\boldsymbol{y}^{N_s})^T]^T\in \mathbb{R}^{N_sN}$ denotes the $N_s$ FIM surface shapes during channel measurement. Therefore, the FIM surface shape optimization problem can be formulated as \vspace{-0.75em}

\begin{subequations}\label{FIM_orthogonal}
\begin{align}
\label{deqn_orthogonal}
\min_{\tilde{\boldsymbol{y}}}\quad &\gamma(\tilde{\boldsymbol{y}})\\
s.t. \quad& \tilde{y}_{\tilde{n}} \in \mathcal{B}, \quad \tilde{n}=1,2,\ldots,N_sN,
\end{align}
\end{subequations}
where $\tilde{y}_{\tilde{n}}$ denotes the $\tilde{n}$-th element in $\tilde{\boldsymbol{y}}$.

Since $\gamma(\tilde{\boldsymbol{y}})$ is highly non-convex with respect to $\tilde{\boldsymbol{y}}$ and has $N_sN$ coupled variables, it is generally difficult to obtain the closed-form solution of problem (\ref{FIM_orthogonal}), albeit its conciseness. To balance convergence speed and the complexity of the optimization algorithm, in this paper, the Davidon-Fletcher-Powell (DFP) method \cite{mamat2018derivative} is used to obtain a suboptimal solution of the FIM surface shape optimization problem in an iterative manner. Specifically, at the $k$-th iteration, the surface shape is updated along the search direction $\boldsymbol{p}^k$, i.e., $\tilde{\boldsymbol{y}}^{k+1}\gets \tilde{\boldsymbol{y}}^k-\tau^k \boldsymbol{p}^k$, where $\tilde{\boldsymbol{y}}^k\in \mathbb{R}^{N_sN}$ and $\tau^k \in \mathbb{R}^+$ denote the FIM surface shape and the step size at the $k$-th iteration, respectively. Considering the FIM morphing range constraint, the updated FIM surface shape should be projected into the feasible set $\mathcal{B}$, i.e.,\vspace{-0.5em}

\begin{equation}
    \tilde{y}_{\tilde{n}}^{k+1} \gets \text{min}\bigl( \text{max}(\tilde{y}_{\tilde{n}}^{k+1}, -b), b \bigl),\quad \tilde{n} = 1,\dots, N_sN.
\end{equation}

According to the DFP method, the search direction $\boldsymbol{p}^k$ is given by\cite{mamat2018derivative}
\begin{align}
\label{deqn_ex16}
\boldsymbol{p}^k=&\boldsymbol{H}^k\nabla \gamma(\tilde{\boldsymbol{y}}^k),\\
\label{deqn_ex15}
\boldsymbol{H}^{k+1}=&\boldsymbol{H}^{k}+\frac{\boldsymbol{\mu}^{k}(\boldsymbol{\mu}^{k})^T}{(\boldsymbol{\mu}^{k})^T \boldsymbol{\nu}^{k}}-\frac{\boldsymbol{H}^{k} \boldsymbol{\nu}^{k}(\boldsymbol{H}^{k} \boldsymbol{\nu}^{k})^T}{(\boldsymbol{H}^{k} \boldsymbol{\nu}^{k})^T\boldsymbol{\nu}^{k}},
\end{align}
where $\boldsymbol{\mu}^{k}=\tilde{\boldsymbol{y}}^{k+1}-\tilde{\boldsymbol{y}}^{k}, \boldsymbol{\nu}^{k}=\nabla \gamma(\tilde{\boldsymbol{y}}^{k+1})-\nabla \gamma(\tilde{\boldsymbol{y}}^k)$. In addition, the gradient of $\gamma(\tilde{\boldsymbol{y}})$ with respect to $\tilde{\boldsymbol{y}}$ is given by $\nabla  \gamma(\tilde{\boldsymbol{y}})=[\frac{\partial \gamma(\tilde{\boldsymbol{y}})}{\partial \tilde{y}_1},\ldots,\frac{\partial \gamma(\tilde{\boldsymbol{y}})}{\partial \tilde{y}_{\tilde{n}}},\ldots,\frac{\partial \gamma(\tilde{\boldsymbol{y}})}{\partial \tilde{y}_{N_sN}}]^T$, where the partial derivative of $\gamma(\tilde{\boldsymbol{y}})$ with respect to $y_{\tilde{n}}$ is represented by
\begin{align}
\label{grad_relate}
\frac{\partial \gamma(\tilde{\boldsymbol{y}})}{\partial \tilde{y}_{\tilde{n}}}
&=\sum_{m=1}^M\sum_{v=m+1}^M 2\pi \lambda \Im\bigl\{ \frac{\boldsymbol{\phi}^H_m \boldsymbol{\phi}_v}{\arrowvert \boldsymbol{\phi}^H_m \boldsymbol{\phi}_v \arrowvert} (\sin{\theta^v}\sin{\phi^v}- \notag\\
&\quad\  \sin{\theta^m}\sin{\phi^m} ) e^{-j\frac{2\pi}{\lambda}({\rho^v_{\tilde{n}}-\rho^m_{\tilde{n}}})}\bigr\}. 
\end{align}\vspace{-1em}

A necessary condition to ensure convergence of the DFP algorithm is that the approximate Hessian matrix is positive definite $\boldsymbol{H}^{k} \succ 0$, which requires that $(\boldsymbol{\mu}^{k})^T\boldsymbol{\nu}^{k}>0$ according to \cite{salih2016partial}. Therefore, the update of the Hessian matrix in (\ref{deqn_ex15}) is performed only when the condition $(\boldsymbol{\mu}^{k})^T\boldsymbol{\nu}^{k}>0$ is satisfied.

\begin{algorithm}[t]
	\renewcommand{\algorithmicrequire}{\textbf{Input:}}
	\renewcommand{\algorithmicensure}{\textbf{Output:}}
        \renewcommand{\algorithmicrepeat}{\textbf{Repeat}}
	\renewcommand{\algorithmicuntil}{\textbf{Until}}
	\caption{FIM Surface Shape Morphing for Minimizing Measurement Matrix Columns' Coherence.}
	\label{alg:alg1}
	\begin{algorithmic}[1]
 \REQUIRE $N, L, N_s, k_{\mathrm{max}}, \mathcal{B}, \delta.$
 \STATE Initialization: $\boldsymbol{H}^0=\boldsymbol{I}_N, k=0.$
 \STATE Randomly generate initial iteration points $\tilde{\boldsymbol{y}}^{0}$ within $\mathcal{B}.$
 \REPEAT
 \STATE \text{Calculate the gradient $\nabla \gamma(\tilde{\boldsymbol{y}}^{k})$ according to (\ref{grad_relate})};
 \STATE \text{Calculate the search direction $\boldsymbol{p}^k$ according to (\ref{deqn_ex16})};
 \STATE \text{Determine the step size $\tau^k$ according to (\ref{leq_1}) and (\ref{leq_2})};
 \STATE \text{Update the FIM's surface shape: $\tilde{\boldsymbol{y}}^{k+1} \gets \tilde{\boldsymbol{y}}^{k}-\tau^k\boldsymbol{p}^k$};
 \STATE \text{Project coordinates into the feasible set $\mathcal{B}$};
 \STATE \text{When $(\boldsymbol{\mu}^{k})^T\boldsymbol{\nu}^{k}>0$ is satisfied, update $\boldsymbol{H}^{k+1}$}
 \text{according to (\ref{deqn_ex15}), otherwise let $\boldsymbol{H}^{k+1}\gets \boldsymbol{H}^{k}$;}
 \STATE \text{Update iteration counter by $k\gets k+1$;}
 \UNTIL {The number of iteration counter $k$ reaches the maximum number $k_{\mathrm{max}}$, or the decrease of $
 \gamma (\tilde{\boldsymbol{y}})$ is negligible, i.e., $\arrowvert \gamma(\tilde{\boldsymbol{y}}^{k})-\gamma(\tilde{\boldsymbol{y}}^{k-1})\arrowvert<\delta$.}
 \ENSURE $\tilde{\boldsymbol{y}}^k$.
	\end{algorithmic} 
\end{algorithm}

Moreover, the Armijo–Goldstein condition is used to select an appropriate step size $\tau^k$, ensuring that the objective function is sufficiently improved in each iteration, i.e., \vspace{-0.25em}
\begin{align} 
\label{leq_1}
\gamma(\tilde{\boldsymbol{y}}^k)+(1-c)\tau^k\nabla \gamma(\tilde{\boldsymbol{y}}^k)^T\boldsymbol{p}^k \leq \gamma(\tilde{\boldsymbol{y}}^k+\tau^k\boldsymbol{p}^k),\\
\label{leq_2}
\gamma(\tilde{\boldsymbol{y}}^k+\tau^k\boldsymbol{p}^k) \leq \gamma(\tilde{\boldsymbol{y}}^k)+c\tau^k\nabla \gamma(\tilde{\boldsymbol{y}}^k)^T\boldsymbol{p}^k, 
\end{align}
where $c\in(0,\frac{1}{2})$ is a control parameter used to adjust the acceptable interval of step size under the Armijo-Goldstein condition\cite{goldstein1967effective}. 

To elaborate, the DFP solution for solving problem (\ref{deqn_orthogonal}) is summarized in \textbf{Algorithm \ref{alg:alg1}}, where $k_{\mathrm{max}}$ denotes the maximum number of iterations. Notably, the combination of the Armijo-Goldstein condition and the DFP algorithm ensures that each iteration of \textbf{Algorithm \ref{alg:alg1}} achieves a sufficient reduction of the objective function in (\ref{deqn_orthogonal}). Second, the column coherence value of the measurement matrix has a lower bound of 0. Based on the above two points, the proposed algorithm for minimizing the column coherence value of the measurement matrix can be guaranteed to be convergent.

In each iteration, the computational complexity of gradient computation is $\mathcal{O}\bigl(N_sNM(M-1)/2\bigr)$, and search direction computation yields complexity of $\mathcal{O}(N_s^2N^2)$. Furthermore, the complexity for updating the step size using the Armijo algorithm is $\mathcal{O}(N_sN)$, while the FIM surface shape updating requires a complexity of $\mathcal{O}(N_sN)$. Additionally, the Hessian matrix computation complexity is $\mathcal{O}(N_s^2N^2)$, and the complexity for calculating $\gamma(\tilde{\boldsymbol{y}}^k)$ is $\mathcal{O}\bigl(M(M-1)/2\bigr)$. Therefore, the total complexity of \textbf{Algorithm \ref{alg:alg1}} is no more than $\mathcal{O}[k_{\rm max}\bigl(2N_s^2N^2+N_sNM(M-1)/2\bigr)]$, with the lower-order terms being removed. Finally, $\tilde{\boldsymbol{y}}^{opt}$ resulting in the lowest $\gamma(\tilde{\boldsymbol{y}})$ is utilized to perform channel estimation with the OMP method over $N_s$ measurements.

\section{FIM Surface Shape Morphing for \\SNR Enhancement}
\label{section:Surface-Shape Morphing}
Additionally, FIM has been proven to effectively enhance the system's SNR by appropriately configuring its surface shape. However, accurate FIM surface-shape morphing relies on perfect channel state information (CSI)\cite{an2025flexibleMISO, an2025flexibleMIMO}. In this section, we utilize the DOA and channel gain estimated in Section \ref{ES-CSI} to perform FIM surface shape optimization and analyze its impact on SNR enhancement based on estimated CSI. Specifically, we consider an FIM-aided downlink communication process similar to that in Fig. \ref{fig_1}. Let $s\in \mathbb{C}$ represent the information signal transmitted from the FIM, and $p$ denotes the transmit power. Therefore, the received signal $f\in \mathbb{C}$ is given by \vspace{-0.75em}

\begin{equation}
f= \sqrt{p}\hat{\boldsymbol{h}}^H(\boldsymbol{y})\boldsymbol{w}s+\eta,
\end{equation}
where $\eta\sim\mathcal{CN}(0,\sigma^2)$ is the AWGN with average noise power $\sigma^2$ at the UE. $\boldsymbol{w}\in \mathbb{C}^{N}$ denotes the transmit beamforming vector at the transmitter. Let $\{\hat{\theta}_l, \hat{\phi}_l, \hat{\xi}_l\}$ denote the estimated CSI, the estimated wireless channel $\hat{\boldsymbol{h}}(\boldsymbol y)$ can be readily constructed based on (\ref{deqn_ex1}). 

For a tentative FIM surface shape configuration $\boldsymbol{y}$, the optimal maximum ratio transmission (MRT) beamformer is $\boldsymbol{w}={\hat{\boldsymbol{h}}(\boldsymbol{y})}/{\Arrowvert \hat{\boldsymbol{h}}(\boldsymbol{y})\Arrowvert_2}.$
Therefore, the SNR at the receiver is given by 
\begin{align}
\label{deqn_SNR}
\text{SNR}=p\frac{\arrowvert \hat{\boldsymbol{h}}^H(\boldsymbol{y})\boldsymbol{w}\arrowvert^2}{\sigma^2}
=p\frac{\hat{\boldsymbol{h}}^H(\boldsymbol{y})\hat{\boldsymbol{h}}(\boldsymbol{y})}{\sigma^2},
\end{align}
which is also a function of the FIM surface shape. In order to maximize the receiver SNR by optimizing $\boldsymbol{y}$ with total transmit power $p$ constraint, the transmission optimization problem can be formulated as 
\begin{subequations}\label{deqn_ex3}
\begin{align}
\max_{\boldsymbol{y}}\quad &a(\boldsymbol{y})=\hat{\boldsymbol{h}}^H(\boldsymbol{y})\hat{\boldsymbol{h}}(\boldsymbol{y})\\
s.t. \quad& y_n \in \mathcal{B}, \quad n=1,2,\ldots,N,
\end{align}
\end{subequations}
and the DFP method in Section \ref{DFP_method} can be adapted to solve this problem. Note that in the objective function of (\ref{deqn_ex3}), the channel gain associated with each element only depends on the corresponding element's position. Therefore, $a(\boldsymbol{\boldsymbol{y}})$ can be expressed as the sum of $N$ independent entries, i.e.,
\begin{align}
a(\boldsymbol{\boldsymbol{y}})=\sum\limits_{k=1}^N \vert \hat{\xi}_1 e^{j\frac{2\pi}{\lambda}\hat{\rho}_k^1}+\ldots+ \hat{\xi}_L e^{j\frac{2\pi}{\lambda}\hat{\rho}_k^L}\vert^2,
\end{align}
and the partial derivative of $a(\boldsymbol{\boldsymbol{y}})$ with respect to different elements' positions can be expressed in a similar form. 

For ease of derivation, the $n$-th row of the estimated $\hat{\boldsymbol{G}}(\boldsymbol{y})$ is represented by $\boldsymbol{u}^T(y_n)=[e^{j\frac{2\pi}{\lambda}\hat{\rho}_n^1},e^{j\frac{2\pi}{\lambda}\hat{\rho}_n^2},\ldots,e^{j\frac{2\pi}{\lambda}\hat{\rho}_n^{L}}] \in \mathbb{C}^{1 \times L}$. Note that the deformation of the $n$-th element only affects the $n$-th entry of channel $\hat{\boldsymbol{h}}(\boldsymbol{y})$, namely $\hat{h}_n(y_n)= \boldsymbol{u}^T(y_n)\hat{\boldsymbol{\xi}}$. Then, the partial derivative of $a(\boldsymbol{y})$ with respect to $y_n$ is given by 
\begin{align} \label{equ:A1}
\frac{\partial a(\boldsymbol{y})}{\partial y_n}
=&\frac{\partial \hat{\boldsymbol{h}}^H(\boldsymbol{y})\hat{\boldsymbol{h}}(\boldsymbol{y})}{\partial y_n}
=\frac{\partial \hat{h}^\ast_n(y_n)\hat{h}_n(y_n)}{\partial y_n} \notag\\
=&-j\frac{2\pi}{\lambda}(\boldsymbol{q}\odot\hat{\boldsymbol{\xi}})^H\boldsymbol{u}^\ast(y_n)\boldsymbol{u}^T(y_n)\hat{\boldsymbol{\xi}}\notag\\
&+j\frac{2\pi}{\lambda}\hat{\boldsymbol{\xi}}^H\boldsymbol{u}^\ast(y_n)\boldsymbol{u}^T(y_n)(\boldsymbol{q}\odot\hat{\boldsymbol{\xi}})\notag\\
=&2\Im\Bigl\{ \frac{2\pi}{\lambda}(\boldsymbol{q}\odot\hat{\boldsymbol{\xi}})^H\boldsymbol{u}^\ast(y_n)\boldsymbol{u}^T(y_n)\hat{\boldsymbol{\xi}} \Bigr\},
\end{align}
where $\boldsymbol{q} \triangleq[\sin{\hat{\theta}^1}\sin{\hat{\phi}^1},\ldots,\sin{\hat{\theta}^{L}}\sin{\hat{\phi}^{L}}]^T \in \mathbb{R}^{L}.$

Then, $\nabla a(\boldsymbol{y})=[\frac{\partial a(\boldsymbol{y})}{\partial y_1},\frac{\partial a(\boldsymbol{y})}{\partial y_2},\ldots,\frac{\partial a(\boldsymbol{y})}{\partial y_N}]^T$ can be given by 

\begin{align} \label{deqn_ex4}
\nabla a(\boldsymbol{y})=&\frac{4\pi}{\lambda}\Im\bigl\{ \bigl(\hat{\boldsymbol{G}}^*(\boldsymbol{y})(\boldsymbol{q}\odot\hat{\boldsymbol{\xi}})^\ast \bigr) \odot \bigl( \hat{\boldsymbol{G}}(\boldsymbol{y})\hat{\boldsymbol{\xi}} \bigr)\bigr\}.
\end{align}

\section{Numerical Results}
In this section, we evaluate the channel estimation performance and the SNR gain of the proposed FIM-aided MISO system. Specifically, an FIM containing $N$ elements is deployed at the BS. The spacing between adjacent elements in the x-axis and z-axis directions are set to half-wavelength, i.e., $d_x=d_z=\lambda/2$. $L$ scatterers are distributed in the wireless environment, and the elevation angle $\theta^l$ and azimuth angle $\phi^l$ of each scatterer are uniformly distributed in $(0,\pi)$. We set average channel power $\gamma^2=\gamma_0^2d^{-\alpha}$, where $\gamma_0^2= -60$ dB denotes the channel gain at the reference distance of 1m, $d=100$ m denotes the propagation distance, and $\alpha$ denotes the path loss exponent, which is set to $\alpha = 2.2$ in this paper. Moreover, we consider communication systems operating at 30 GHz with a bandwidth of 50 MHz, and noise spectral density of $-174\ \rm{dBm/Hz}$.

For the channel estimation, we set the number of meta-atoms to $N= 25$, and the morphing boundary $b= \lambda$. The number of observation slots is $N_s= 10$, and the number of columns in the measurement matrix $\boldsymbol{\Psi}$ is $M_x \times M_z= 20 \times 20=400$. To evaluate the channel estimation performance, we use the normalized minimum mean square error (NMSE) as a performance metric, i.e., $\text{NMSE}=\mathbb{E}\bigl[\Arrowvert\boldsymbol{h}-\hat{\boldsymbol{h}}\Arrowvert_2^2 /\Arrowvert\boldsymbol{h}\Arrowvert_2^2\bigr]$, where $\hat{\boldsymbol{h}}$ denotes the uplink channel reconstructed by the estimated DOA and multipath gain. Additionally, a rigid UPA array or RIS having the same number of elements with FIM and half-wavelength antenna spacing is considered as a benchmark scheme. For \textbf{Algorithm \ref{alg:alg1}}, the control parameter in (\ref{leq_1}) is set to $c=0.33$, while the maximum number of iterations and the iteration stop condition are $k_{\mathrm{max}}=50$ and $\delta=10^{-4}$, respectively. Considering that the result of a single DFP optimization may yield a local optimal solution, we set multiple random initial iteration points in \textbf{Algorithm \ref{alg:alg1}} to improve the performance of the optimization algorithm. The number of random initial points is set to 10. Besides, all experimental results are obtained by averaging 100 Monte-Carlo simulations.



\begin{figure*}[!t]
	\centering
	\begin{minipage}{0.32\linewidth}
		\centering
		\includegraphics[width=1.1\linewidth]{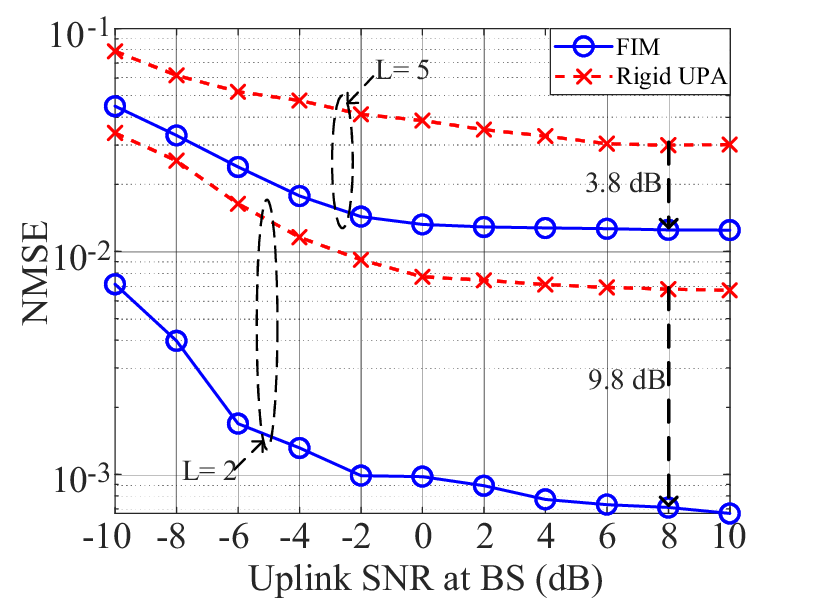}
		\caption{NMSE of channel estimates versus the uplink SNR.}
		\label{fig_uplink_SNR}
	\end{minipage}
	\begin{minipage}{0.32\linewidth}
		\centering
		\includegraphics[width=1.1\linewidth]{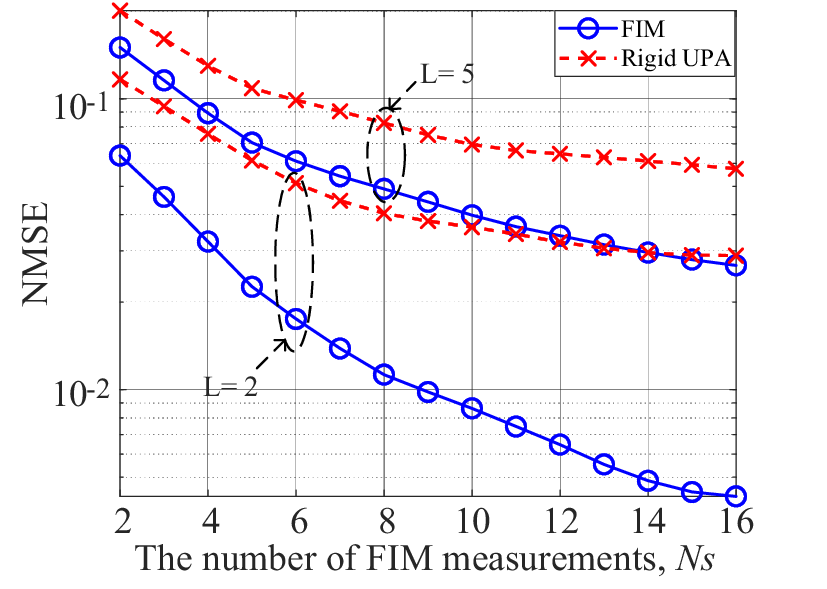}
		\caption{NMSE of channel estimates versus the number of measurements.}
		\label{fig_measurement}
	\end{minipage}
	\begin{minipage}{0.32\linewidth}
		\centering
		\includegraphics[width=1.05\linewidth]{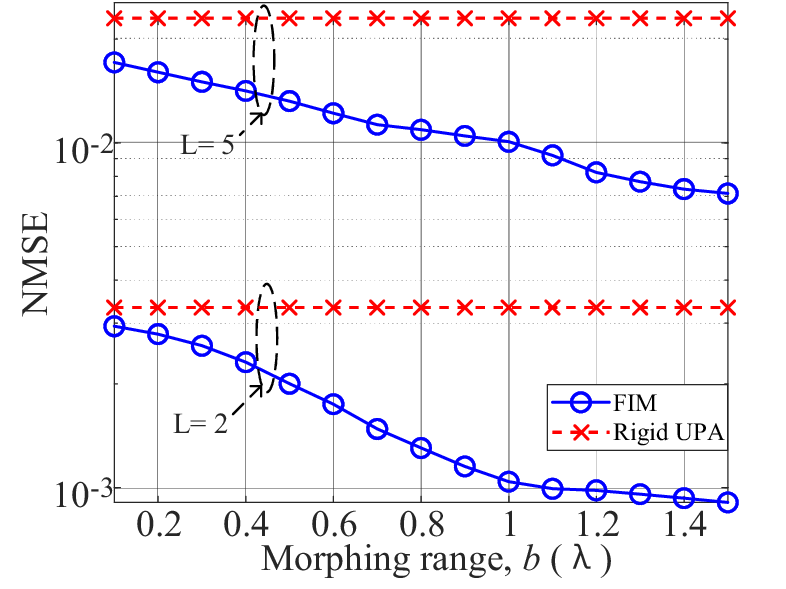}
		\caption{NMSE of channel estimates versus the morphing range.}
		\label{fig_morphing_range}
	\end{minipage}
\end{figure*}

\begin{figure*}[!t]
	\centering
	    \begin{minipage}{0.31\linewidth}
		\centering
		\includegraphics[width=1.05\linewidth]{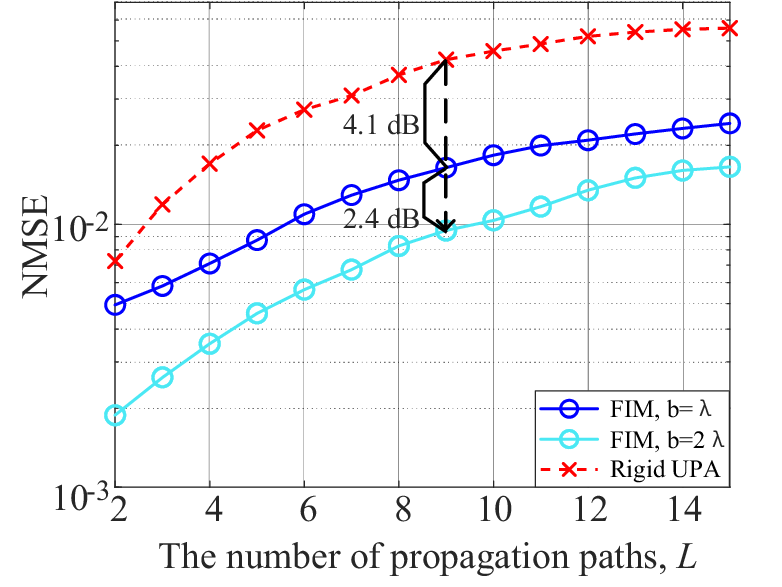}
		\caption{NMSE of channel estimates versus the number of paths.}
		\label{fig_path}
	\end{minipage}
	\begin{minipage}{0.31\linewidth}
		\centering
		\includegraphics[width=1.1\linewidth]{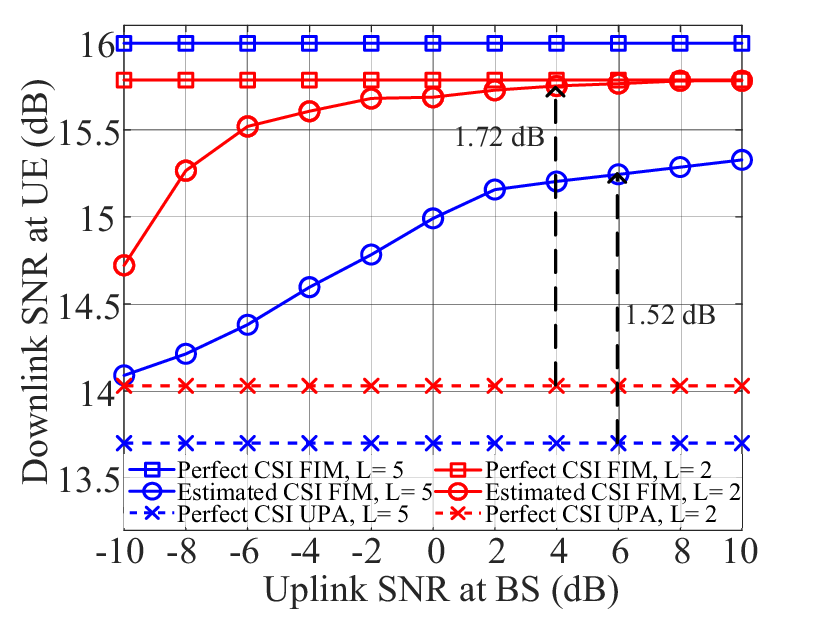}
		\caption{The downlink SNR gain at the UE based on the estimated CSI.}
		\label{fig_SNR_gain}
	\end{minipage}
	\begin{minipage}{0.33\linewidth}
		\centering
		\includegraphics[width=1.1\linewidth]{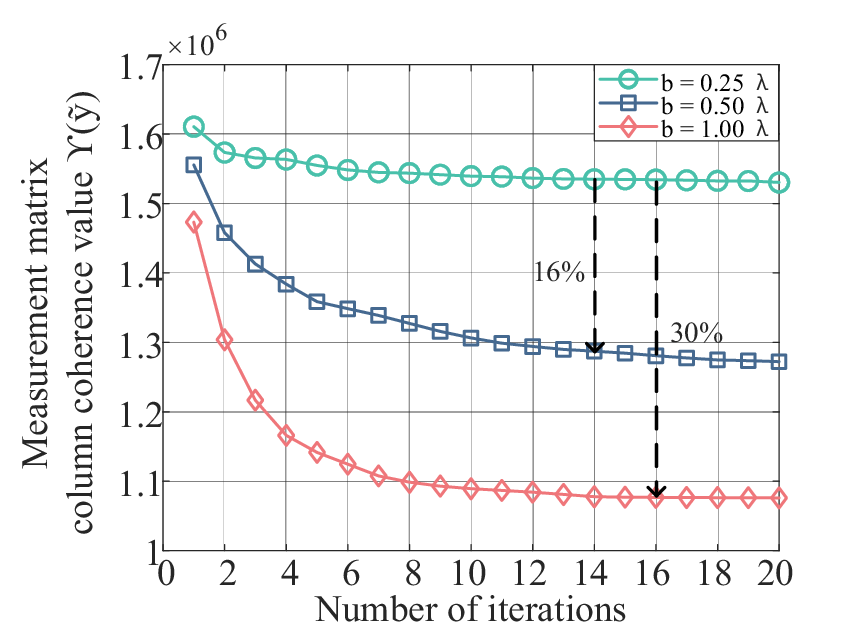}
		\caption{Column coherence value of measurement matrix versus the number of iterations.}
		\label{fig_DFP}
	\end{minipage}
\end{figure*}

We observe from Fig. \ref{fig_uplink_SNR} that the increase of uplink SNR will reduce the estimation error. Furthermore, when compared with rigid UPA, the FIM can provide a prominent gain for improving the channel estimation performance by morphing its surface shape over multiple time slots. Specifically, FIM decreases the NMSE by $9.8$ dB when considering $L=2$ and uplink SNR $= 8$ dB. Moreover, as the number of scatterers in the environment increases, the channel estimation performance of both FIM and UPA degrades to some extent. Nonetheless, FIM can still achieve a 3.8 dB improvement in channel estimation NMSE when $L = 5$.

In Fig. \ref{fig_measurement}, we analyze the impact of the number of channel measurements $N_s$ on the NMSE, considering uplink SNR $ = -10$ dB. Fig. \ref{fig_measurement} demonstrates that increasing $N_s$ or reducing the number of scatterers can further improve the performance of the compressed sensing-based channel estimator. This is because multiple measurements help mitigate the impact of noise on channel estimation and provide greater flexibility for FIM to enhance the orthogonality of the measurement matrix.

In Fig. \ref{fig_morphing_range}, we analyze the impact of morphing range on the channel estimation performance, considering uplink SNR of $10$ dB. As expected, a larger morphing range contributes to the improvement of FIM's channel estimation performance by expanding the optimization space for designing the measurement matrix, which demonstrates the advantage of FIM in enhancing channel estimation performance. Nonetheless, a diminishing return of NMSE reduction is observed.

Fig. \ref{fig_path} evaluates the NMSE versus the number of propagation paths, considering $N_s = 10$ channel measurements and uplink SNR $ = 10$ dB. Across all experimental setups, an increase in the number of propagation paths in the environment leads to a rise in channel estimation NMSE. Nevertheless, the channel estimation performance by using an FIM remains superior to that using conventional UPAs, even in complex communication environments. Specifically, when considering $L=9$, the FIM with morphing range of $b=\lambda$ achieves a $4.1$ dB reduction in NMSE compared to the rigid UPA. Additionally, expanding the FIM deformation range can improve the FIM’s channel estimation performance to a certain extent.

Finally, Fig. \ref{fig_SNR_gain} shows the SNR gain at UE of the FIM-enhanced downlink system using the estimated CSI. For comparison, we also use the perfect CSI to morph the surface shape of the transmit FIM to maximize the SNR gain. The downlink transmit signal power is set as $p=3$ dBm. It can be observed that the incorporation of FIM effectively enhances the SNR at UE. Moreover, a higher uplink SNR during channel estimation yields a lower channel estimation error, leading to a higher SNR for UE in the subsequent downlink communication stage. However, due to the impact of channel estimation errors, FIM suffers from moderate gain loss compared to that based on the optimal surface shape. Nonetheless, substantial SNR gain can be offered compared to the conventional rigid UPA. Specifically, when considering $L=5$, the FIM still achieves a 1.72 dB SNR improvement over UPA. Notably, when $L=2$, the FIM designed based on estimated channel parameters can almost achieve the performance gain obtained under perfect CSI conditions. 

To evaluate the convergence of \textbf{Algorithm \ref{alg:alg1}}, we show the details of the column coherence value of the measurement matrix during the iterative processes. As shown in Fig. \ref{fig_DFP}, under all three settings of $b$, the column coherence value exhibits a trend of initially decreasing and then stabilizing, converging rapidly within about 10 iterations. Moreover, a larger value of $b$ brings more freedom for optimization, which leads to a lower column coherence value when stability is reached, which means that \textbf{Algorithm \ref{alg:alg1}} can benefit from a larger FIM morphing range.

\section{Conclusion}
In this paper, we utilized an FIM to enhance the channel estimation performance in a single-user communication system. Specifically, we first utilized the OMP method to formulate the channel estimation as a sparse recovery problem to estimate the channel parameters in FIM-aided uplink communication systems. Then, the DFP algorithm was utilized to morph the FIM surface shape to decrease the column coherence of the measurement matrix. Additionally, the DFP algorithm was also used to optimize the FIM surface shape to improve the received SNR of the FIM-aided downlink communication system based on the estimated DOA and channel gain. Numerical results validated the performance advantage of FIM over rigid UPA in enhancing channel estimation accuracy and communication SNR. In particular, when considering scenarios with fewer scatterers, FIM can even achieve performance gain close to that under perfect CSI conditions. The application of FIM in fast time-varying channels or high-handover scenarios requires further investigation. Meanwhile, we will consider more complex communication scenarios, such as multi-user MIMO communication systems. Additionally, given the deformation characteristics of FIM, it becomes crucial to characterize the performance gain under the practical constraints and energy consumption issues related to the deployment of FIM.

\bibliographystyle{IEEEtran}
\bibliography{references}

@ARTICLE{10158690,
  author={An, Jiancheng and Xu, Chao and Ng, Derrick Wing Kwan and Alexandropoulos, George C. and Huang, Chongwen and Yuen, Chau and Hanzo, Lajos},
  journal={IEEE J. Sel. Areas Commun.}, 
  title={Stacked Intelligent Metasurfaces for Efficient Holographic {MIMO} Communications in {6G}}, 
  year={2023},
  month={Aug.},
  volume={41},
  number={8},
  pages={2380-2396},
  doi={10.1109/JSAC.2023.3288261}
}

@ARTICLE{9040264,
  author={Giordani, Marco and Polese, Michele and Mezzavilla, Marco and Rangan, Sundeep and Zorzi, Michele},
  journal={IEEE Commun. Mag.}, 
  title={Toward {6G} Networks: Use Cases and Technologies}, 
  year={2020},
  month={Mar.},
  volume={58},
  number={3},
  pages={55-61},
  doi={10.1109/MCOM.001.1900411}
}

@article{zheng2023flexible,
  title={Flexible-position {MIMO} for wireless communications: Fundamentals, challenges, and future directions},
  author={Zheng, Jiakang and Zhang, Jiayi and Du, Hongyang and Niyato, Dusit and Sun, Sumei and Ai, Bo and Letaief, Khaled B},
  journal={IEEE Wireless Communications},
  year={2024},
  month={Oct.},
  volume={31},
  number={5},
  pages={18-26},
  publisher={IEEE}
}

@article{ni2021dynamically,
  title={A dynamically reprogrammable surface with self-evolving shape morphing},
  author={Bai, Yun and Wang, Heling and Xue, Yeguang and Pan, Yuxin and Kim, Jin-Tae and Ni, Xinchen and Liu, Tzu-Li and Yang, Yiyuan and Han, Mengdi and Huang, Yonggang and others},
  journal={Nature},
  volume={609},
  number={7928},
  pages={701--708},
  year={2022},
  month={Dec.},
  publisher={Nature Publishing Group UK London}
}

@article{zhang2023self,
  title={Self-healing cellulose-based flexible sensor: A review},
  author={Zhang, Yue-hong and Lei, Qin-yang and Liu, Rui-jing and Zhang, Lei and Lyu, Bin and Liu, Lei-peng and Ma, Jian-zhong},
  journal={Industrial Crops and Products},
  volume={206},
  pages={117724},
  year={2023},
  month={Dec.},
  publisher={Elsevier}
}

@article{niu2021reconfigurable,
  title={Reconfigurable shape-morphing flexible surfaces realized by individually addressable photoactuator arrays},
  author={Niu, Dong and Jiang, Weitao and Li, Dachao and Ye, Guoyong and Luo, Feng and Liu, Hongzhong},
  journal={Smart Materials and Structures},
  volume={30},
  number={12},
  pages={125032},
  year={2021},
  month={Nov.},
  publisher={IOP Publishing}
}

@inproceedings{mamat2018derivative,
  title={Derivative free {Davidon-Fletcher-Powell} {(DFP)} for solving symmetric systems of nonlinear equations},
  author={Mamat, M and Dauda, MK and Mohamed, MA bin and Waziri, MY and Mohamad, FS and Abdullah, H},
  booktitle={IOP Conference Series: Materials Science and Engineering},
  volume={332},
  year={2018}
}

@article{heath2016overview,
  title={An overview of signal processing techniques for millimeter wave {MIMO} systems},
  author={Heath, Robert W and Gonzalez-Prelcic, Nuria and Rangan, Sundeep and Roh, Wonil and Sayeed, Akbar M},
  journal={IEEE J. Sel. Topics in Signal Process.},
  volume={10},
  number={3},
  pages={436--453},
  year={2016},
  month={Feb.},
  publisher={IEEE}
}

@ARTICLE{8240645,
  author={Shahmansoori, Arash and Garcia, Gabriel E. and Destino, Giuseppe and Seco-Granados, Gonzalo and Wymeersch, Henk},
  journal={IEEE Trans. Wireless Commun.}, 
  title={Position and Orientation Estimation Through Millimeter-Wave {MIMO} in {5G} Systems}, 
  year={2018},
  month={Mar.},
  volume={17},
  number={3},
  pages={1822-1835},
  doi={10.1109/TWC.2017.2785788}}

@ARTICLE{salih2016partial,
  title={Partial Davidon, Fletcher and Powell {(DFP)} of quasi newton method for unconstrained optimization},
  author={Salih, Basheer M and Abbo, Khalil K and Abdullah, Zeyad M},
  journal={Tikrit J. Pure Sc},
  volume={21},
  number={6},
  pages={180--186},
  year={2016}
}

@article{an2024emerging,
  title={Emerging Technologies in Intelligent Metasurfaces: Shaping the Future of Wireless Communications},
  author={An, Jiancheng and Debbah, M{\'e}rouane and Cui, Tie Jun and Chen, Zhi Ning and Yuen, Chau},
  journal={arXiv preprint arXiv:2411.19754},
  year={2024}
}

@article{goldstein1967effective,
  title={An effective algorithm for minimization},
  author={Goldstein, AA and Price, JF},
  journal={Numer. Math.},
  volume={10},
  pages={184--189},
  year={1967},
  month={Oct.},
  publisher={Springer}
}

@ARTICLE{WC_2024_An_Codebook,
  author={An, Jiancheng and Xu, Chao and Wu, Qingqing and Ng, Derrick Wing Kwan and Di Renzo, Marco and Yuen, Chau and Hanzo, Lajos},
  journal={IEEE Wireless Commun.}, 
  title={Codebook-Based Solutions for Reconfigurable Intelligent Surfaces and Their Open Challenges}, 
  year={2024},
  month={Apr.},
  volume={31},
  number={2},
  pages={134-141},
  doi={10.1109/MWC.010.2200312}
}

@ARTICLE{TCOM_2024_Yu_Environment,
  author={Yu, Zhiheng and An, Jiancheng and Basar, Ertugrul and Gan, Lu and Yuen, Chau},
  journal={IEEE Trans. Commun.}, 
  title={Environment-Aware Codebook Design for {RIS}-Assisted {MU-MISO} Communications: Implementation and Performance Analysis}, 
  year={2024},
  month={Dec.},
  volume={72},
  number={12},
  pages={7466-7479},
  doi={10.1109/TCOMM.2024.3415619}
}

@article{zhang2024target,
  title={Target Detection and Positioning Aided by Reconfigurable Surfaces: Reflective or Holographic?},
  author={Zhang, Xiaoyu and Zhang, Haobo and Liu, Liang and Han, Zhu and Poor, H Vincent and Di, Boya},
  journal={IEEE Trans. Wireless Commun.},
  year={2024},
  month={Dec.},
  volume={23},
  number={12},
  pages={19215-19230},
  publisher={IEEE}
}

@article{yuan2025reconfigurable,
  title={Reconfigurable Holographic Surface Enhanced Multi-User Terrestrial-Satellite Communications},
  author={Yuan, Chenyang and Yu, Gang and Xu, Sai and Liu, Yiliang and Zhang, Jie},
  journal={IEEE Trans. Veh. Technol.},
  year={2025},
  month={Mar.},
  volume={},
  number={},
  pages={1-6},
  publisher={IEEE}
}

@article{an2025flexibleMISO,
  title={Flexible intelligent metasurfaces for downlink multiuser {MISO} communications},
  author={An, Jiancheng and Yuen, Chau and Di Renzo, Marco and Debbah, M{\'e}rouane and Poor, H Vincent and Hanzo, Lajos},
  journal={IEEE Trans. Wireless Commun.},
  year={2025},
  month={Jan.},
  volume={24},
  number={4},
  pages={2940-2955},
  publisher={IEEE}
}

@article{an2025flexibleMIMO,
  title={Flexible intelligent metasurfaces for enhancing {MIMO} communications},
  author={An, Jiancheng and Han, Zhu and Niyato, Dusit and Debbah, M{\'e}rouane and Yuen, Chau and Hanzo, Lajos},
  journal={IEEE Trans. Commun.},
  year={2025},
  month={Mar.},
  volume={},
  number={},
  pages={1-1},
  publisher={IEEE}
}

@article{FIM_folding,
  title     = {Programmable kiri-kirigami metamaterials},
  author    = {Tang, Yichao and Lin, Gaojian and Yang, Shu and Yi, Yun Kyu and Kamien, Randall D and Yin, Jie},
  journal   = {Adv. Mater.},
  volume    = {29},
  number    = {10},
  pages     = {1604262},
  year      = {2017},
  publisher = {Wiley Online Library}
}

@article{FIM_stretchable,
  title     = {Pangolin-Inspired Stretchable, Microwave-Invisible Metascale},
  author    = {Wang, Changxian and Lv, Zhisheng and Mohan, Manoj Prabhakar and Cui, Zequn and Liu, Zhihua and Jiang, Ying and Li, Jiaofu and Wang, Cong and Pan, Shaowu and Karim, Muhammad Faeyz and others},
  journal   = {Adv. Mater.},
  volume    = {33},
  number    = {41},
  pages     = {2102131},
  year      = {2021},
  publisher = {Wiley Online Library}
}

@article{jiang2023bivariate,
  title={Bivariate pilot optimization for compressed channel estimation in {RIS}-assisted multiuser {MISO-OFDM} systems},
  author={Jiang, Rongkun and Fei, Zesong and Huang, Shihan and Wang, Xinyi and Wu, Qingqing and Ren, Shiwei},
  journal={IEEE Transactions on Vehicular Technology},
  volume={72},
  number={7},
  pages={9115--9130},
  year={2023},
  publisher={IEEE}
}
\end{document}